# Assessing Cyclostationary Malware Detection via Feature Selection and Classification

Mike Nkongolo[1][0000−0003−0938−113X]

University of Pretoria, Department of Informatics, Faculty of Engineering, Built
Environment and Information Technology
mike.wankongolo@up.ac.za

**Abstract.** Cyclostationarity involves periodic statistical variations in
signals and processes, commonly used in signal analysis and network se-
curity. In the context of attacks, cyclostationarity helps detect malicious
behaviors within network traffic, such as traffic patterns in Distributed
Denial of Service (DDoS) attacks or hidden communication channels in
malware. This approach enhances security by identifying abnormal pat-
terns and informing Network Intrusion Detection Systems (NIDSs) to
recognize potential attacks, enhancing protection against both known
and novel threats. This research focuses on identifying cyclostationary
malware behavior and its detection. The main goal is to pinpoint es-
sential cyclostationary features used in NIDSs. These features are ex-
tracted using algorithms such as Boruta and Principal Component Anal-
ysis (PCA), and then categorized to find the most significant cyclosta-
tionary patterns. The aim of this article is to reveal periodically changing
malware behaviors through cyclostationarity. The study highlights the
importance of spotting cyclostationary malware in NIDSs by using es-
tablished datasets like KDD99, NSL-KDD, and the UGRansome dataset.
The UGRansome dataset is designed for anomaly detection research and
includes both normal and abnormal network threat categories of zero-
day attacks. A comparison is made using the Random Forest (RF) and
Support Vector Machine (SVM) algorithms, while also evaluating the
effectiveness of Boruta and PCA. The findings show that PCA is more
promising than using Boruta alone for extracting cyclostationary net-
work feature patterns. Additionally, the analysis identifies the internet
protocol as the most noticeable cyclostationary feature pattern used by
malware. Notably, the UGRansome dataset outperforms the KDD99 and
NSL-KDD, achieving 99% accuracy in signature malware detection using
the RF algorithm and 98% with the SVM. The research suggests that
the UGRansome dataset is a valuable choice for studying anomaly and
cyclostationary malware detection efficiently. Lastly, the study recom-
mends using the RF algorithm for effectively categorizing and detecting
cyclostationary malware behaviors.

tion, signature malware detection, UGRansome.

# 1   Introduction

The domain of Network Intrusion Detection Systems (NIDS) seeks innovative strategies to detect anomalous traffic patterns that surpass conventional malware detection methods [18, 15]. Recognizing cyclostationary traffic patterns holds significant potential to enhance NIDS efficiency and facilitate the implementation of pioneering frameworks [12]. Most current NIDS solutions overlook the use of cyclostationary techniques [11, 19] for pattern detection, which could differentiate static from dynamic patterns [20]. Detecting cyclostationary traffic patterns aids in discerning if an intrusion manifests as a long-term evolving malware, undergoing periodic changes [14]. This study aims to evaluate the practices of long-term evolving malware from a cyclostationary perspective. The term cyclostationary is employed to characterize the traffic patterns of zero-day threats [17], which can vary based on network attributes. This research analyzes the cyclostationarity of both known and unknown zero-day threats [16], driven by the dearth of cyclostationary datasets for Network Intrusion Detection Problems (NIDP) to comprehend the cyclostationarity of long-term evolving malware like zero-day threats. In contrast to previous methodologies, this article avoids unnecessary abstraction of cyclostationarity. It addresses the omission of evaluating cyclostationary traffic patterns for zero-day threat detection, a gap prevalent in previous NIDS research. Previous studies on cyclostationarity of zero-day threat behaviors primarily adopt anomaly detection techniques, which are the most comprehensible. The concept of cyclostationarity finds application in various scientific and engineering fields [2, 9, 8, 7, 10]. For instance, Mechanical Engineering employs periodicity and cyclostationarity to analyze the behavior of rotating and reciprocating components. Meteorology studies cyclostationarity in relation to weather prediction owing to Earth's revolution and rotation influencing seasonal variations. In Network Communication, disciplines like radar, telemetry, and sonar capitalize on periodicity and cyclostationarity, essential for signal scanning, sampling, multiplexing, and modulation [7]. The article's experiments provide diverse insights for the NIDP, as limited efforts concern cyclostationarity within the Network Intrusion Detection Landscape (NIDL). The primary contribution lies in utilizing Boruta, PCA, and Supervised Learning algorithms to detect cyclostationarity in zero-day threats. The article's structure encompasses theoretical explication of cyclostationarity and related works in Section 2, followed by the proposed methodology and datasets for detecting cyclostationary zero-day threats in Section 3. Experimental results concerning cyclostationary feature pattern recognition are presented in Section 4, while Section 5 furnishes the research's conclusions.

## 2   Literature review

### 2.1   Description of a cyclostationary malware: How it gets differ over a traditional malware

A cyclostationary malware represents a type of network threat characterized by its irregular attributes that exhibit cyclic variations over time [13]. In the context of cyclostationary network traffic, the traffic can be divided into discrete segments such as $T_1, T_2, ..., T_n$, with the malwares often concealed within these anomalous traffic segments. Detecting these malwares necessitates the identification of the abnormal sample, which can then be further broken down into contiguous or separate segments. These segments are subsequently leveraged for classification and in-depth analysis of the detected cyclostationary malwares. Within each segment, individual data points signify distinct cyclostationary feature patterns of specific zero-day threats. To elaborate further, the term traditional malwares refers to network attacks that have become obsolete and are no longer commonly used. In contrast, cyclostationary malwares employ updated protocols that are integrated into the Transmission Control Protocol (TCP) suite. The distinguishing factor lies in their protocols—while cyclostationary malwares are aligned with modern TCP/IP suite protocols, traditional malwares rely on outdated protocols that have fallen out of usage. This transition underscores the evolving nature of cyber threats, with cyclostationary malwares exploiting contemporary protocols for their malicious activities.

### 2.2   Related works

Network traffic, whether in the long-term or short-term, often exhibits periodic behaviors. The periodic nature of malware presents an effective feature for Network Intrusion Detection Systems (NIDSs) design and performance assessment. Employing anomaly detection enables the establishment of thresholds for identifying anomalies or recognizing cyclostationary threats based on network flow volume. However, these threshold values are subject to variation over time. Recent research by Yinka et al. [29] employed threshold values to distinguish cyclostationary network traffic from stationary patterns. Their classification process demonstrated enhanced evaluation metrics, but the evolving thresholds remain a challenge. In another study [22], ensemble learning evaluated cyclostationarity using heterogeneous datasets. The stacking technique incorporated a feature selection method to efficiently detect relevant features and accurately identify cyclostationary traffic in the network. Vivekanandam et al. [28] proposed an adaptable Machine Learning approach involving a genetic algorithm for feature selection. This method, combined with other algorithms, demonstrated improved performance in detecting diverse malware categories. Similarly, Mugunthan et al. [5] introduced a cloud-based architecture using a Markov Model and the Random Forest algorithm to detect malwares in network flows, especially low-level Distributed Denial of Service (DDoS) attacks. The rise of long-term evolution malware, including ransomware, emphasizes cyclostationarity as a primary source of intrusion by zero-day threats. Lin et al. [4] reported that, while long-term evolution devices constitute only a small percentage of internet connectivity, they contribute significantly to cyclostationary network traffic. As users transition from PCs to mobile devices, hackers exploit mobile device vulnerabilities, driving the proliferation of zero-day threats. Analyzing the cyclostationarity of zero-day threats via network traffic pattern deviations becomes crucial. In a different context, Raja et al. [21] explored the implications of pervasive computing in electric motors, considering correlations between objects, infrastructure, and urban expansion. This highlights the evolving landscape of computing environments. In general, research underscores the importance of cyclostationarity in NIDS, demonstrating its potential in improving intrusion detection and response mechanisms while addressing the challenges posed by evolving threat landscapes and shifting device preferences.

## 2.3   Limitations in existing works

The NIDL has a solid background in terms of normal threats detection methodologies but lacks the analysis of stochastic, cyclostationary traffic, queuing of network flow, intrusion modelisation, and zero-day threats taxonomy [1]. Figure 1 shows the framework used to assess cyclostationarity of malwares.

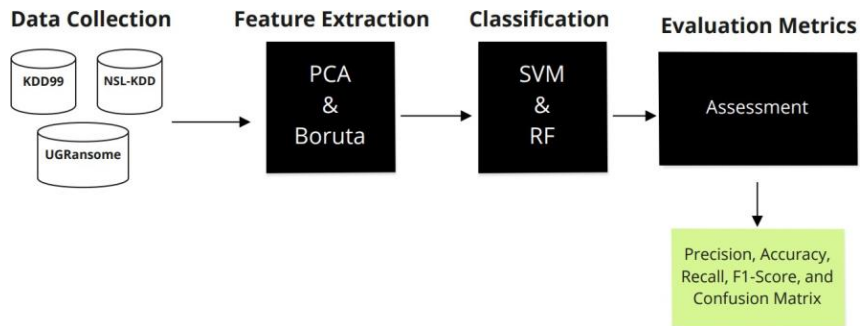

**Fig. 1.** An overview of the supervised model for cyclostationary malware detection.

The application of cyclostationary analysis in studying network traffic patterns remains an underutilized approach within the Network Intrusion Detection Problem (NIDP) domain. To address this untapped potential, we introduce a supervised Machine Learning model [26], previously employed across diverse NIDPs (depicted in Figure 1). This Supervised Learning framework forms the basis for cyclostationary malware detection, with a focus on legacy datasets like the Knowledge Discovery and Data Mining (KDD99) and Network Security Laboratory–Knowledge Discovery and Data Mining (NSL-KDD), as illustrated in Figure 1. Our objective is to uncover cyclostationary patterns within these datasets, paralleled by the application of the cyclostationary dataset, UGRansome [13], to achieve the same goal. The Supervised Learning framework employs two key algorithms, the Support Vector Machine (SVM) and Random Forest (RF), selected for comparative analysis. Through evaluation metrics such as Confusion Matrix, Recall, F1-Score, Precision, and Accuracy, we measure the performance of these algorithms. Notably, certain methodologies to assess the cyclostationarity of malware evolution may require specialized skills. Unlike conventional techniques focused on detecting normal attacks, recognizing the unique attributes of long-term evolving malware like zero-day threats demands periodicity detection. This intricate process relies heavily on time and often necessitates rare or transient process analysis. In this context, the supervised approach to cyclostationary malware detection presents a valuable tool for designing and implementing Network Intrusion Detection Systems (NIDSs) geared toward the detection of zero-day threats. By harnessing supervised learning techniques, we facilitate the incorporation of cyclostationarity as a discriminative factor in NIDSs, thereby enhancing their sensitivity and effectiveness in tackling the evolving landscape of network security threats.

## 2.4   The KDD99 dataset

The KDD Cup 99 dataset, as depicted in Figure 2, was initially established as a benchmark data source for Network Intrusion Detection Systems (NIDSs), evaluated at MIT's Lincoln Lab and sponsored by DARPA between 1998 and 1999. This dataset encompasses five predictive categories, including Remote to Local (R2L), Probe, User to Root (U2R), and Denial of Service (DoS), serving to categorize diverse network threats [24]. Moreover, even the normal behaviors of different malwares are included in this dataset. It comprises 41 attributes classified into Traffic, Content, and Basic categories. The majority of network threats fall within the DoS and Normal classes, with a proportion of 98.6% [24]. As highlighted in Figure 2, the imbalanced nature of this dataset becomes apparent. This imbalance signifies a scenario where one class is more prevalent than another. Consequently, the data distribution tilts in favor of a specific category, potentially biasing the Machine Learning classification outcomes towards that favored class [13]. The training set of the KDD99 dataset contains 4,898,431 rows, corresponding to 2,984,154 observations. The duplicate features are present in both the testing and training subsets [24]. However, it is important to note that the KDD99 dataset, being outdated, might not be ideally suited for cyclostationarity analysis, as indicated in Figure 2. To provide further insight, Table 1 offers a comprehensive overview of the malware instances identified in the KDD99 and NSL-KDD datasets. This examination of the KDD Cup 99 dataset underscores the complexities and considerations tied to real-world network intrusion detection scenarios. As technology evolves, datasets designed for earlier purposes might not seamlessly align with contemporary analysis requirements, highlighting the need for updated and contextually relevant datasets in the study of network security [13, 17].

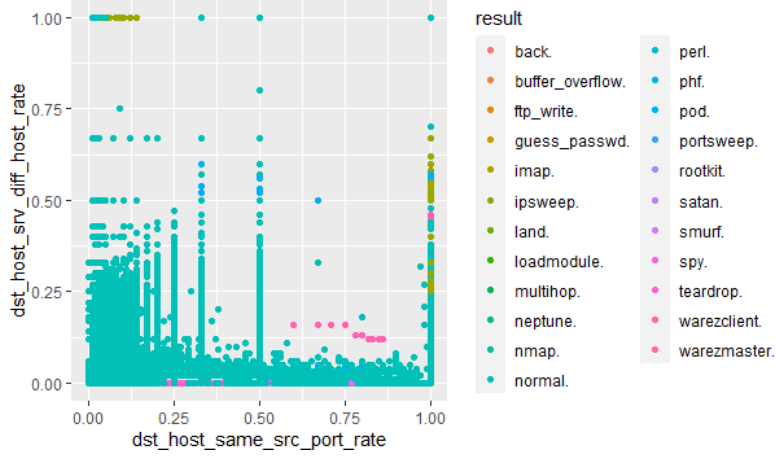

**Fig. 2.** Imbalanced network threats of the KDD99 dataset. Normal attacks are more represented compared to other attacks.

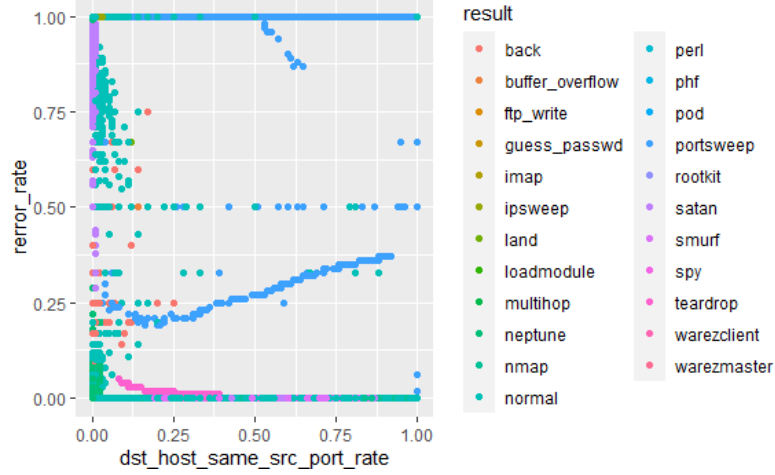

**Fig. 3.** The normalised NSL-KDD dataset. The proportion of normal threats is more or less balanced.

## 2.5 The NSL- KDD dataset

The KDD Cup 99 dataset, as illustrated in Figure 2, emerged as a seminal benchmark for evaluating Network Intrusion Detection Systems (NIDSs) during its period of assessment at MIT's Lincoln Lab, sponsored by DARPA from 1998 to 1999. This dataset encompasses five predictive categories - Remote to Local (R2L), Probe, User to Root (U2R), and Denial of Service (DoS) - serving as classification criteria for diverse network threats [24]. Interestingly, it also includes normal behaviors of various malwares, presenting a comprehensive perspective. The inherent imbalance in these classes, as evidenced in Figure 2, is a critical observation. This imbalance signifies a scenario where one class is disproportionately prevalent compared to another, potentially introducing bias in Machine Learning classification outcomes towards the overrepresented class [13]. However, it is essential to consider that while the KDD99 dataset served as a foundational resource, its obsolescence may impact its suitability for contemporary analysis, particularly in the context of cyclostationarity, as indicated by Figure 2. This limitation arises from the dataset's age and the evolving nature of network threats and behaviors. In light of this, the applicability of the KDD99 dataset to the study of cyclostationarity in malwares is constrained by its design for a different era of network security challenges. As technological landscapes transform, it becomes crucial to align datasets with the specific requirements of modern intrusion detection methodologies, including the emerging focus on cyclostationary analysis in detecting network threats. This underscores the dynamic nature of network security research and the continual need for relevant and up-to-date data sources to effectively address contemporary cybersecurity concerns.

**Table 1.** Malware detected in the KDD99 and NSL-KDD datasets.

| DoS | Probe | U2R | R2L |
|---|---|---|---|
| Back | IpSweep | Buffer Overflow | FtpWrite |
| Land | Nmap | LoadModule | GuessPassword |
| Neptune | PortSweep | Perl | Imap |
| Smurf | Satan | RootKit | Multihop |
| Teardrop | NA | NA | phf, WarezMaster, Spy |

## 3 Methodology

The methodological approach to classifying cyclostationary malware through Supervised Learning is orchestrated in two overarching phases, each encapsulating distinct objectives within the framework delineated in Figure 1. The initial phase encompasses the extraction of pivotal features intrinsic to the cyclostationary context, whereas the ensuing phase involves their systematic classification, thus conferring a structured taxonomy upon the delineated features. Our proposed methodology is scaffolded upon the employment of two pivotal algorithms, Boruta and Principal Component Analysis (PCA), which serve as the bedrock for feature extraction. While PCA conventionally operates as a dimensionality reduction tool, its role as a feature extraction conduit is justified by its inherent capability to both unveil the pertinence of data components and unveil their variance. The crucible of this methodology resides in the strategic amalgamation of these feature extractor components, poised to elicit intrinsic patterns from the datasets of interest. Once the vital features are distilled from the aforementioned datasets, the mantle of classification is assumed by the proficiency of the Random Forest and Support Vector Machine algorithms, entrusted with the task of meticulous stratification. These algorithms traverse the multidimensional feature space, forging connections and discerning relationships, ultimately ascribing individual instances to their rightful cyclostationary niches. The zenith of this methodology culminates in the meticulous evaluation of the adopted algorithms, specifically the Random Forest and Support Vector Machine. This phase affords insight into the algorithms' efficacy and performance within the context of cyclostationary malware classification. The outcomes borne of this evaluation, replete with nuances and insights, are subsequently laid bare for comprehensive scrutiny and discourse. Figure 1 stands as a visual embodiment of the intricate choreography of steps requisite for the harmonious implementation of the proposed methodology. The journey commences with the meticulous collection of pertinent data, the lifeblood of this endeavor, and traverses through the analytical labyrinth, culminating in the pivotal Supervised Learning evaluation phase expounded upon in the forthcoming sections. In alignment with experimental rigor, the dataset was subject to rigorous cross-validation, an empirical stratagem where the dataset is partitioned into training and testing sets - an 80% allocation for training and the remaining 20% for comprehensive testing. This rigorous validation methodology serves as a robust safeguard against overfitting and ensures the integrity of results obtained through this comprehensive process.

## 3.1    The cyclostationary dataset

The experimental landscape was enriched by the incorporation of the UGRansome dataset, a potent asset in our pursuit (Figure 4). This dataset emerges as a synthesis of the UGR'16 and ransomware datasets, a union yielding 207,534 distinct cyclostationary features and 14 attributes within 14 tuples [13]. These meticulously triangulated features have been harnessed through a Data Fusion technique, conferring upon them the strategic potency to serve as adept instruments for both anomaly detection and the identification of cyclostationary zero-day threats. A pivotal attribute of the UGRansome dataset is its nuanced stratification of various long-term malware instances into a tripartite predictive class configuration - Signature (S), Synthetic Signature (SS), and Anomaly (A) - outlined succinctly in Figure 4. Furthermore, this dataset boasts a granular classification of 16 distinct ransomware families, encompassing entities such as advanced persistent threats (APT), Locky, DMALocker, SamSam, and more, all meticulously clustered and correlated for efficient computational maneuvering [13]. The UGRansome dataset, a product of meticulous construction in the year 2021 [13], stands as a publicly accessible resource. Its core distinguishing facet rests in its inherently cyclostationary and periodic nature, epitomizing a rich, real-world portrayal of network traffic dynamics. This character becomes vividly apparent in Figure 5, where anomalous malware-induced network traffic exhibits a variance linked to distinct network flags. An elegant equilibrium characterizes the distribution of novel malware instances, adeptly balanced across more than 100,000 Internet Protocol addresses, their assignments distributed coherently across class A, B, C, and D classifications. The versatility of the UGRansome dataset transcends its immediate realm of cyclostationary zero-day threat identification. Its expansive potential extends to facilitating the discernment of other malware archetypes, including SSH, Bonet, DoS, Port Scanning, NerisBonet, and Scan, widening the horizons of its applicability [13]. To provide comprehensive insight, Table 2 elucidates the intricate data structure intrinsic to the UGRansome dataset, an instrumental aid in comprehending its manifold dimensions.

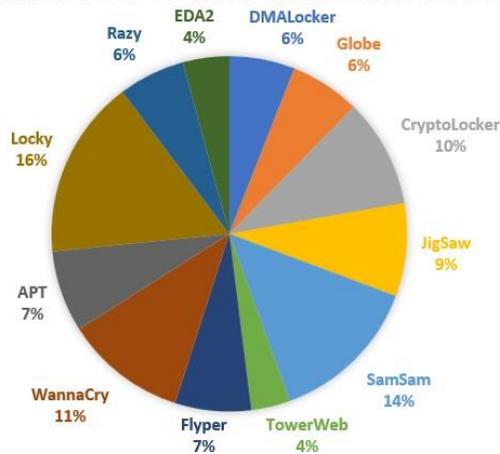

**Fig. 4.** The malware classes of the cylostationary dataset.

### 3.2   Feature extraction with Boruta

In the realm of feature extraction, the efficacy of Boruta has been harnessed to gauge the significance of features inherent to the KDD99 and NSL-KDD datasets. Boruta stands as a potent algorithmic feature extractor, leveraging the prowess of the Random Forest methodology to tackle an array of regression and classification challenges [3, 6]. Rooted in the foundations of Decision Trees, Boruta operates via a multi-faceted approach, independently cultivating various Decision Trees across diverse samples extracted from the training corpus. This algorithm adopts a Wrapper technique, augmenting the original dataset by introducing shadow features, entities imbued with random values meticulously permuted among training observations [3, 23]. The crux of Boruta's prowess is unveiled through its evaluation of feature relevance. This assessment unfolds through the lens of stratification accuracy, a pivotal metric in the context of supervised classification tasks [3, 6, 23]. The orchestration of Boruta's relevance computation unfolds in two successive steps. Initially, the stratification accuracy loss is evaluated individually within the canopy of Decision Trees, facilitating a nuanced evaluation where each Tree may assign disparate classifications to a given feature. Subsequently, the amalgamation of average and Standard Deviation computations bestows the essence of the stratification accuracy loss with comprehensible quantitative measures. The Z-score assumes center stage as the linchpin in Boruta's calculus of feature importance [3, 6, 23]. Computed by standardizing the average loss by the Standard Deviation, the Z-score serves as a conduit for evaluating and juxtaposing feature relevance. This measured relevance cascades into a tiered categorization of features, bifurcating their utility into three discernible strata:

1. Rejected: Features that exhibit minimal import and are consequently cast aside.
2. Confirmed: Features of pronounced import, substantiated by their high Z-scores, thus meriting their inclusion.

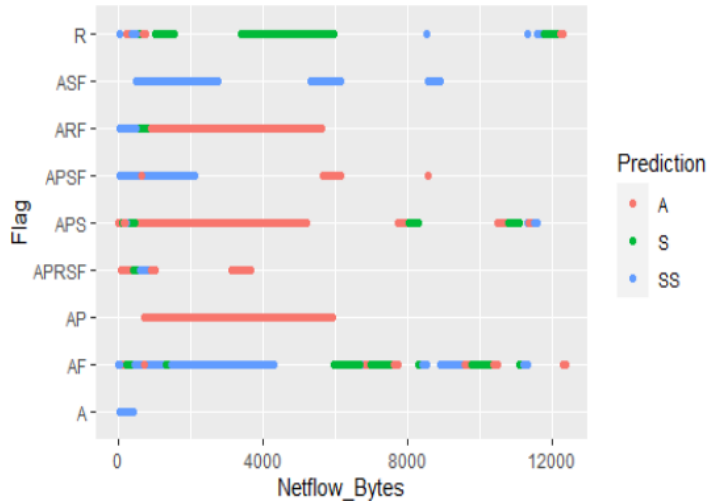

**Fig. 5.** The cyclostationarity of anomalous network traffic patterns.

**Table 2.** The structure of the UGRansome dataset.

| Number | Attribute | Example | Description | Cyclostationarity |
|--------|-----------|---------|-------------|-------------------|
| 1 | Prediction | SS | Synthetic Signature malware | Yes |
| 2 | Ransomware | WannaCry | Novel malware | Yes |
| 3 | Bitcoins (BTC) | 60.0 BTC | Ransom payment | Yes |
| 4 | Dollars (USD) | 400 USD | Ransom payment | Yes |
| 5 | Cluster | 1 | Group assigned per malware | Yes |
| 6 | Seed Address | 1dice6yg | Malware address | Yes |
| 7 | Expended Address | 4ePEyKtk | Malware address | Yes |
| 8 | Port | 5062 | Communication endpoint | Yes |
| 9 | Malware | Bonet | Novel malware | Yes |
| 10 | Network traffic | 1819 000 | Periodic network flow | Yes |
| 11 | IP address | Class A | Unique address identifying a device | Yes |
| 12 | Flag | AF | Network state | Yes |
| 13 | Protocol | TCP | Communication rule | Yes |
| 14 | Timestamp | 40 seconds | Netflow termination | Yes |

3. Tentative: Features that dwell in the grey area, characterized by uncertain Z-scores, thereby warranting further scrutiny.

In pursuit of this end, the maximization of the Z-score stands as a critical juncture. The Maximum Z-Score (MZS) observed across shadow features becomes the benchmark, against which the Z-scores of primary features are juxtaposed [3, 6, 23]. Features outshining the MZS are embraced in the realm of confirmed relevance, while those trailing behind are relegated to the rejected category. This methodological orchestration ensures that Boruta's scrutiny bestows upon the feature selection process an optimal mix of precision and comprehensiveness.

### 3.3   Feature extraction with PCA

Principal Component Analysis (PCA) is a mathematical technique used for reducing the complexity of data while preserving its essential patterns. It transforms a set of correlated variables into a new set of uncorrelated variables called principal components. These components capture the most significant variations in the data, allowing for efficient visualization, dimensionality reduction, and feature extraction [27]. PCA is commonly applied in various fields, including data analysis, image processing, and machine learning, to reveal underlying structures and simplify data representation. In the context of Principal Component Analysis (PCA), the recorded features within the $i^{th}$ category are encapsulated by the notation $y_{k,l}$. These features are organized into $j^{th}$ instances and represented as elements of an $n \times p$ matrix denoted as $Y$. Prior to any analysis, it is imperative to bring the dataset into a standardized form, ensuring each column adheres to a distribution characterized by a zero mean and Standard Deviation. It is within this context that the pivotal PCA normalization process takes place, en-riching the features for further exploration and analysis [27]. The normalization procedure is mathematically expressed as follows:

$$x_{k,l} = \frac{y_{k,l} - \hat{y}_l}{s_l},$$  (1)

In this equation, the terms $s_l$ and $\hat{y}_l$ represent the Standard Deviation and mean of the corresponding column within matrix $Y$. More explicitly:

– $y_{k,l}$ signifies the value of the $k^{th}$ instance of the $l^{th}$ feature in matrix $Y$.
– $\hat{y}_l$ stands for the mean of the $l^{th}$ feature across all instances.
– $s_l$ symbolizes the Standard Deviation of the $l^{th}$ feature across all instances.

By undertaking this normalization, each feature is rendered dimensionless, removing any inherent biases stemming from variations in scales or units. Consequently, PCA can robustly capture the underlying patterns and variations in the dataset, leading to the extraction of principal components that succinctly encapsulate its most salient characteristics. These normalized features become the foundation for the generation of principal components, facilitating effective dimensionality reduction and aiding in tasks such as feature selection and anomaly detection. The transformative power of normalization within the PCA framework extends beyond mathematical manipulation; it is a methodical way to prepare data for a journey of insightful discovery within a lower-dimensional space.

### 3.4   Random Forest algorithm

This algorithm constructs individualistic Decision Trees from the training sample. Predictions are pooled from all Trees to make the final result of classification. In short, the Random Forest algorithm utilises a set of results to make a final prediction/ classification, and they are commonly named Ensemble Learning approaches [25]. The relevance of features is computed by using the decrease in the impurity of weighted nodes. The probability is computed by using the frequency of features in the node, subdivided by the sum of all samples [25]. The greatest value represents the most important feature in the dataset. The total of feature's relevance value is computed and subdivided by the number of Trees:

$$RF = \frac{\sum\limits_{l \in T} N_{k,l}}{T},$$  (2)

where the Random Forest is denoted by RF with normalised features relevance $N_{k,l}$ and $T$ is the number of Trees.

### 3.5   Support Vector Machine algorithm

The Support Vector Machine sorts features into binary or multiple categories by using a threshold as a separative measure (Figure 6). Each feature is represented by a data point in the hyperplane and the Lagrange formula is generally computed to segregate different categories. Lastly, the Euclidean distance is calculated between the threshold and data points to draw a boundary that distinguishes clusters (Figure 6). The boundary differentiating data points can be written as follows:

$$H : W^T(x) + b = 0,\tag{3}$$

W is the weighted features while $x$ denotes original inputs features. The hyperplane is denoted by $H$ and its bias by $b$.

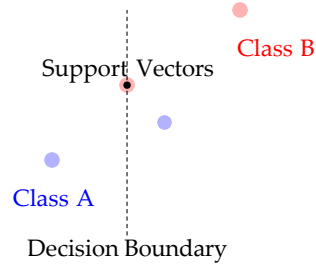

**Fig. 6.** Support Vector Machine (SVM) Decision Boundary

### 3.6   Evaluation and testing of the Supervised Learning framework

The following evaluation metrics are used to assess the proposed framework. False Positive (FP) and False Negative (FN) represent misclassification while the correct classification is represented by True Positive (TP) and True Negative (TN):

$$Accuracy = \frac{TN + TP}{TN + TP + FP + FN},\tag{4}$$

this metric represents the ratio of accurate classification for both TN (True Negative) and True Positive (TP) cases. The computational time is used to assess the feature extraction performance while the Confusion Matrix evaluates the Supervised Learning algorithms by tabulating the correct stratification results.

$$Precision = \frac{TP}{TP + FP},\tag{5}$$

The Recall value specifies the actual positive cases predicted correctly by the Machine Learning algorithm. The formula is as follows:

$$Recall = \frac{TP}{TP + FN},\tag{6}$$

The F1-Score converges the mean of Recall and Precision and show the overall combined evaluation performance:

$$F1 - Score = \frac{2 * Precision * Recall}{Precision + Recall}, \qquad (7)$$

The Confusion Matrix is a $n * n$ matrix used in the computation of TP, TN, FN, and FP to calculate the evaluation metrics such as Accuracy, Precision, and Recall. In this research, the results of the classification using the aforementioned evaluation metrics will be tabulated. Lastly, the computational framework is tested with cross-validation. The computing environment representing the hardware and software specification framework is illustrated in Table 3. The random state of 42 is used for cross-validation.

**Table 3.** Hardware and Software Specifications

| Node | Type |
|------|------|
| Test set | 20% |
| Train set | 80% |
| Random state | 42 |
| Classifier | RF & SVM |
| Feature extraction | Boruta & PCA |
| Number of trees | 100 |
| Dataset | KDD99, NSL-KDD99, and UGRansome |
| Processor | 2.59 GHz |
| System | 64-bit |
| Language | R with RStudio |
| RAM | 39 GB |
| Operating System | Windows |
| Computer | Lenovo |

## 4   Results

The experiment involved training the three datasets using carefully selected Supervised Learning algorithms within the Rstudio computing environment. The caret library was harnessed to utilize Machine Learning packages on a 64-bit Windows 10 Operating System. The Boruta algorithm was employed on the KDD99 and NSL-KDD datasets, as detailed in Table 4. It is worth mentioning that Boruta took relatively more computational time for the NSL-KDD dataset and required more iterations, leading to the rejection of a greater number of features compared to the KDD99 and UGRansome datasets. The outcomes of the PCA algorithm, applied to the UGRansome dataset, were visually illustrated in Figure 7, showcasing the recognition of network protocol (TCP) as the predominant cyclostationary feature pattern, with an occurrence of 92,157 instances. Further insights were drawn from Figure **??**, which highlights the exemplary performance of the Random Forest and Support Vector Machine algorithms, achieving an impressive 99% Accuracy on the UGRansome dataset (Figure 8). The culmination of the experiment was encapsulated by the Confusion Matrix presented in Figure 9, which assessed the Random Forest algorithm's performance. The UGRansome dataset exhibited remarkable results compared to the KDD99 and NSL-KDD datasets in effectively categorizing cyclostationary feature patterns into three distinct attack categories: Signature (S), Synthetic Signature (SS), and Anomaly (A) (Figure 9). Specifically, the detection of Signature malware was particularly prominent, occurring 17,891 times. In summary, this computational exploration underscores the viability of PCA for extracting and classifying cyclostationary network feature patterns. The preeminent cyclostationary feature pattern pertains to the network protocol. Moreover, the UGRansome dataset exhibited superior performance in detecting signature malware when compared to the KDD99 and NSL-KDD datasets.

**Table 4.** Boruta Results using KDD99 and NSL-KDD Datasets

| Dataset | Time | Iterations | Important | Rejected |
|---|---|---|---|---|
| KDD99 | 1.2 hours | 12 | 36 | 2 |
| NSL-KDD | 23092 hours | 24 | 36 | 3 |
| UGRansome | 30 minutes | 4 | 3 | 0 |

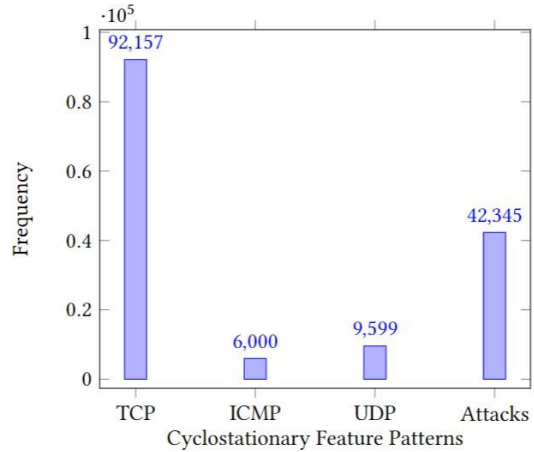

**Fig. 7.** The UGRansome classification results

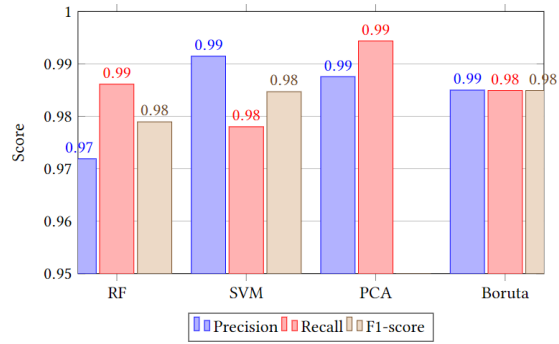

**Fig. 8.** The overall classification results of the UGRansome data

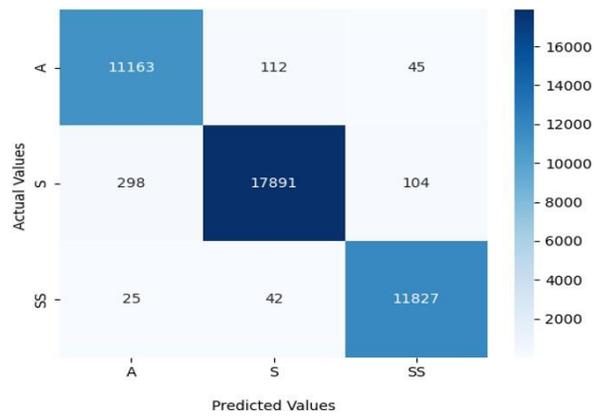

**Fig. 9.** The UGRansome confusion matrix

## 5  Conclusion

In cybersecurity, the pressing challenge of identifying elusive zero-day attacks characterized by cyclostationary behaviors necessitates the deployment of sophisticated methods. In response, this research endeavor delved into the intricate landscape of cyclostationarity using a diverse triad of datasets. The core objective was to decipher the cyclostationary nature of long-term evolution malware, and to this end, a feature extraction paradigm, bolstered by the synergistic prowess of Boruta and Random Forest algorithms, was devised. The focal datasets, namely KDD99, NSL-KDD, and UGRansome, were meticulously scrutinized to unearth latent cyclostationary patterns. The acquired insights were not confined to extraction alone; they extended to the realm of classification. Leveraging the robust capabilities of the Random Forest and Support Vector Machine algorithms, the cyclostationary features were seamlessly classified. This meticulous classification yielded remarkable outcomes—specifically, an outstanding 98% Accuracy on the NSL-KDD dataset and an impressive 99% on both the KDD99 and UGRansome datasets. These findings stand as a testament to the potency of intelligent algorithms in accurately detecting cyclostationary patterns in evolving malware scenarios. Yet, as the landscape of cyber threats continues to evolve, so must our methodologies. While this study has significantly illuminated the path toward understanding cyclostationarity in malware behavior, it also underscores the need for more comprehensive approaches. The current experiment predominantly thrives within the realms of Supervised Learning, prompting an imperative exploration into the realm of Deep Learning for a more nuanced and agile analysis of cyclostationarity in long-term evolution malware. As the horizon of research expands, one intriguing avenue remains unexplored: the evaluation of feature extraction efficacy via the prism of a Genetic Algorithm applied to the UGRansome dataset. Such an exploration could potentially unravel latent insights, further refining our arsenal against the persistent threat of cyclostationary malware. In the continued pursuit of securing digital landscapes, the research community must forge ahead with a multidisciplinary approach, harnessing the power of intelligent algorithms and cutting-edge methodologies to fortify our defenses against the ever-evolving cyber frontier.

## References

[1] Abed Saif Alghawli. "Complex methods detect anomalies in real time based on time series analysis". In: *Alexandria Engineering Journal* 61.1 (2022), pp. 549–561.

[2] Venessa Darwin and Mike Nkongolo. "Data Protection for Data Privacy-A South African Problem?" In: *arXiv preprint arXiv:2306.09934* (2023).

[3] Alif Nur Iman and Tohari Ahmad. "Improving intrusion detection system by estimating parameters of random forest in Boruta". In: *2020 International Conference on Smart Technology and Applications (ICoSTA)*. IEEE. 2020, pp. 1–6.

[4] Cheng Yuan Lin, BaiHua Chen, and WeiYao Lan. "An Efficient Approach for Encrypted Traffic Classification using CNN and Bidirectional GRU". In: *2022 2nd International Conference on Consumer Electronics and Computer Engineering (ICCECE)*. IEEE. 2022, pp. 368–373.

[5] SR Mugunthan. "Soft computing based autonomous low rate DDOS attack detection and security for cloud computing". In: *J. Soft Comput. Paradig.(JSCP)* 1.02 (2019), pp. 80–90.

[6] Divya Nehra, Veenu Mangat, and Krishan Kumar. "A Deep Learning Approach for Network Intrusion Detection Using Non-symmetric Autoencoder". In: *Intelligent Computing and Communication Systems*. Springer, 2021, pp. 371–382. doi: https://doi.org/10.1007/978-981-16-1295-4_38.

[7] Tshimankinda Jerome Ngoy and Mike Nkongolo. "Software-based signal compression algorithm for ROM-stored electrical cables". In: *arXiv preprint arXiv:2308.11620* (2023).

[8] Mike Nkongolo. "Fuzzification-based Feature Selection for Enhanced Website Content Encryption". In: *arXiv preprint arXiv:2306.13548* (2023).

[9] Mike Nkongolo. "Fuzzy feature selection with key-based cryptographic transformations". In: *arXiv preprint arXiv:2306.09583* (2023).

[10] Mike Nkongolo. "Navigating the complex nexus: cybersecurity in political landscapes". In: *arXiv preprint arXiv:2308.08005* (2023).

[11] Mike Nkongolo. "Using ARIMA to Predict the Growth in the Subscriber Data Usage". In: *Eng* 4.1 (2023), pp. 92–120.

[12] Mike Nkongolo, Jacobus Philippus van Deventer, and Sydney Mambwe Kasongo. "The application of cyclostationary malware detection using boruta and pca". In: *Computer Networks and Inventive Communication Technologies: Proceedings of Fifth ICCNCT 2022*. Springer, 2022, pp. 547–562.

[13] Mike Nkongolo, Jacobus Philippus van Deventer, and Sydney Mambwe Kasongo. "UGRansome1819: A Novel Dataset for Anomaly Detection and Zero-Day Threats". In: *Information* 12.10 (2021), p. 405.

[14] Mike Nkongolo, Jacobus Phillipus van Deventer, and Sydney Mambwe Kasongo. "Using deep packet inspection data to examine subscribers on the network". In: *Procedia Computer Science* 215 (2022), pp. 182–191.

[15] Mike Nkongolo, Nita Mennega, and Izaan van Zyl. "Cybersecurity Career Requirements: A Literature Review". In: *arXiv preprint arXiv:2306.09599* (2023).

[16] Mike Nkongolo and Mahmut Tokmak. "Zero-Day Threats Detection for Critical Infrastructures". In: *South African Institute of Computer Scientists and Information Technologists*. Ed. by Aurona Gerber and Marijke Coetzee. Cham: Springer Nature Switzerland, 2023, pp. 32–47. isbn: 978-3-031-39652-6.

[17] Mike Nkongolo and Mahmut Tokmak. "Zero-day threats detection for critical infrastructures". In: *arXiv preprint arXiv:2306.06366* (2023).

[18] Mike Nkongolo et al. "A cloud based optimization method for zero-day threats detection using genetic algorithm and ensemble learning". In: *Electronics* 11.11 (2022), p. 1749.

[19] Mike Nkongolo et al. "Classifying social media using deep packet inspection data". In: *Inventive Communication and Computational Technologies: Proceedings of ICICCT 2022*. Springer, 2022, pp. 543–557.

[20] Mike Nkongolo et al. "Network policy enforcement: An intrusion prevention approach for critical infrastructures". In: *2022 6th International Conference on Electronics, Communication and Aerospace Technology*. IEEE. 2022, pp. 686–692.

[21] K Raja and M Lilly Florence. "Implementation of IDS Within a Crew Using ID3Algorithm in Wireless Sensor Local Area Network". In: *International Conference on Inventive Computation Technologies*. Springer. 2019, pp. 467–475.

[22] Mamunur Rashid et al. "A tree-based stacking ensemble technique with feature selection for network intrusion detection". In: *Applied Intelligence* (2022), pp. 1–14.

[23] Azar Abid Salih and Adnan Mohsin Abdulazeez. "Evaluation of classification algorithms for intrusion detection system: A review". In: *Journal of Soft Computing and Data Mining* 2.1 (2021), pp. 31–40.

[24] Ch Sekhar et al. "Deep Learning Algorithms for Intrusion Detection Systems: Extensive Comparison Analysis". In: *Turkish Journal of Computer and Mathematics Education (TURCOMAT)* 12.11 (2021), pp. 2990–3000.

[25] Sandeep Shah et al. "Implementing a network intrusion detection system using semi-supervised support vector machine and random forest". In: *Proceedings of the 2021 ACM Southeast Conference*. 2021, pp. 180–184. doi: https://doi.org/10.1145/3409334.3452073.

[26] Ahsen Tahir et al. "Hrnn4f: Hybrid deep random neural network for multichannel fall activity detection". In: *Probability in the Engineering and Informational Sciences* 35.1 (2021), pp. 37–50.

[27] Insha Ullah et al. "Detection of cybersecurity attacks through analysis of web browsing activities using principal component analysis". In: *arXiv preprint arXiv:2107.12592* (2021).

[28] B Vivekanandam. "Design an Adaptive Hybrid Approach for Genetic Algorithm to Detect Effective Malware Detection in Android Division". In: *Journal of Ubiquitous Computing and Communication Technologies* 3.2 (2021), pp. 135–149.

[29] Chika Yinka-Banjo et al. "Intrusion Detection Using Anomaly Detection Algorithm and Snort". In: *Illumination of Artificial Intelligence in Cybersecurity and Forensics*. Springer, 2022, pp. 45–70.